\documentstyle[prd,aps,preprint,tighten,epsfig]{revtex}

\begin{document}

\draft

\title{Casas-Ibarra Parametrization and Unflavored Leptogenesis}
\author{{\bf Zhi-zhong Xing}
\thanks{E-mail: xingzz@ihep.ac.cn}}
\address{Institute of High Energy Physics
and Theoretical Physics Center for Science Facilities, \\ Chinese
Academy of Sciences, P.O. Box 918, Beijing 100049, China}
\maketitle

\begin{abstract}
The Casas-Ibarra parametrization is a description of the Dirac
neutrino mass matrix $M^{}_{\rm D}$ in terms of the neutrino mixing
matrix $V$, an orthogonal matrix $O$ and the diagonal mass matrices
of light and heavy Majorana neutrinos in the type-I seesaw
mechanism. Because $M^\dagger_{\rm D} M^{}_{\rm D}$ is apparently
independent of $V$ but dependent on $O$ in this parametrization, a
number of authors have claimed that unflavored leptogenesis has
nothing to do with CP violation at low energies. Here we question
this logic by clarifying the physical meaning of $O$. We establish a
clear relationship between $O$ and the observable quantities, and
find that $O$ {\it does} depend on $V$. We show that both unflavored
leptogenesis and flavored leptogenesis have no direct connection
with low-energy CP violation.
\end{abstract}


\newpage

\framebox{\large\bf 1} ~ Very compelling evidence for finite
neutrino masses and large neutrino mixing angles has been achieved
from solar \cite{SNO}, atmospheric \cite{SK}, reactor \cite{KM}
and accelerator \cite{K2K} neutrino oscillation experiments. This
exciting breakthrough opens a new window to physics beyond the
standard electroweak model, because the standard model itself only
contains three massless neutrinos whose flavor states
$\nu^{}_\alpha$ (for $\alpha = e, \mu, \tau$) and mass states
$\nu^{}_i$ (for $i=1, 2, 3$) are identical. A very natural and
elegant way of generating non-zero but tiny masses $m^{}_i$ for
$\nu^{}_i$ is to extend the standard model by introducing three
right-handed neutrinos and allowing lepton number violation. In
this case, the $SU(2)_{\rm L} \times U(1)_{\rm Y}$ gauge-invariant
mass terms of charged leptons and neutrinos are given by
\begin{equation}
-{\cal L}^{}_{\rm mass} \; =\; \overline{l^{}_{\rm L}} Y^{}_l H
E^{}_{\rm R} + \overline{l^{}_{\rm L}} Y^{}_\nu \tilde{H}
N^{}_{\rm R} + \frac{1}{2} \overline{N^{c}_{\rm R}} M^{}_{\rm R}
N^{}_{\rm R} + {\rm h.c.} \; ,
\end{equation}
where $\tilde{H} \equiv i\sigma^{~}_2 H^*$, $l_{\rm L}$ denotes
the left-handed lepton doublet, and $M^{}_{\rm R}$ is the mass
matrix of right-handed neutrinos. After spontaneous gauge symmetry
breaking, we are left with the charged-lepton mass matrix $M^{}_l
= Y^{}_l v$ and the Dirac neutrino mass matrix $M^{}_{\rm D} =
Y^{}_\nu v$, where $v \simeq 174 ~ {\rm GeV}$ is the vacuum
expectation value of the neutral component of the Higgs doublet
$H$. The scale of $M^{}_{\rm R}$ can be much higher than $v$, as
right-handed neutrinos belong to the $SU(2)^{}_{\rm L}$ singlet
and are not subject to electroweak symmetry breaking. It is
therefore natural to obtain the effective mass matrix for three
light neutrinos \cite{SS1}:
\begin{equation}
M^{}_\nu \; \approx \; -M^{}_{\rm D} M^{-1}_{\rm R} M^T_{\rm D} \;
.
\end{equation}
Such a relation is commonly referred to as the type-I seesaw
mechanism. Let us denote the mass states of three right-handed
neutrinos and their corresponding masses as $N^{}_i$ and $M^{}_i$
(for $i=1,2,3$), respectively. Then Eq. (2) implies $m^{}_i \sim
v^2/M^{}_i$ as a naive result, which explains why $m^{}_i$ is
small but non-vanishing. Note that both light and heavy neutrinos
are Majorana particles in this seesaw picture. Without loss of
generality, one usually chooses the basis with both $Y^{}_l$ (or
$M^{}_l$) and $M^{}_{\rm R}$ being diagonal, real and positive
(i.e., $M^{}_l = {\rm Diag}\{m^{}_e, m^{}_\mu, m^{}_\tau\}$ and
$M^{}_{\rm R} = \widehat{M}^{}_N \equiv {\rm Diag}\{M^{}_1,
M^{}_2, M^{}_3\}$). In this basis, Casas and Ibarra (CI) proposed
an interesting parametrization of $M^{}_{\rm D}$ \cite{CI}:
\begin{equation}
M^{}_{\rm D} \; \approx \; i V \sqrt{\widehat{M}^{}_\nu} ~ O
\sqrt{\widehat{M}^{}_N} \;\; ,
\end{equation}
where $V$ is the $3\times 3$ neutrino mixing matrix which can be
obtained from the diagonalization of $M^{}_\nu$ (i.e., $V^\dagger
M^{}_\nu V^* = \widehat{M}^{}_\nu \equiv {\rm Diag}\{m^{}_1,
m^{}_2, m^{}_3\}$)
\footnote{Note that we have tentatively ignored tiny differences
between the eigenvalues of $M^{}_{\rm R}$ (or $M^{}_\nu$) and the
physical masses $M^{}_i$ (or $m^{}_i$). See the next section for a
detailed discussion.},
and $O$ is a complex orthogonal matrix.

Associated with the above seesaw mechanism, the leptogenesis
mechanism \cite{FY} may naturally work to account for the
cosmological matter-antimatter asymmetry via the CP-violating and
out-of-equilibrium decays of $N^{}_i$ and the $(B-L)$-conserving
sphaleron processes \cite{Kuzmin}. The CP-violating asymmetry
between $N^{}_i \rightarrow l + H^c$ and $N^{}_i \rightarrow l +
H$ decays, denoted as $\varepsilon^{}_i$ (for $i=1,2,3$), has been
calculated in the so-called single flavor approximation (i.e., the
final-state lepton flavors are not distinguished and are simply
summed) \cite{UF}:
\begin{equation}
\varepsilon^{}_i \; = \; \frac{1}{8\pi v^2 (M^\dagger_{\rm D}
M^{}_{\rm D})^{}_{ii}} \sum_{j\neq i} \left\{ {\cal F}(x^{}_{ij}) ~
{\rm Im}\left[ (M^\dagger_{\rm D} M^{}_{\rm D})^{}_{ij} \right]^2
\right\} \; ,
\end{equation}
where ${\cal F}(x^{}_{ij}) = \sqrt{x^{}_{ij}}
\{(2-x^{}_{ij})/(1-x^{}_{ij}) + (1+x^{}_{ij}) \ln
[x^{}_{ij}/(1+x^{}_{ij})]\}$ with $x^{}_{ij} \equiv M^2_j/M^2_i$
is the loop function of self-energy and vertex corrections. In
this {\it unflavored} leptogenesis scenario, a non-vanishing
$\varepsilon^{}_i$ depends on the imaginary part of
$M^\dagger_{\rm D} M^{}_{\rm D}$. Given the CI parametrization in
Eq. (3), it is straightforward to obtain
\begin{equation}
M^\dagger_{\rm D} M^{}_{\rm D} \; \approx \;
\sqrt{\widehat{M}^{}_N} ~ O^\dagger \widehat{M}^{}_\nu O
\sqrt{\widehat{M}^{}_N} \;\; ,
\end{equation}
which is apparently independent of $V$ but dependent on $O$. Hence a
number of authors have taken it for granted that unflavored
leptogenesis has nothing to do with CP violation at low energies
(see, e.g., Refs. \cite{Masina,Rebelo,Petcov,Buras,Petcov2,Branco}).
We find that this conclusion is questionable, because the physical
meaning of $O$ has never been clarified in the literature.

The main purpose of this note is to clarify the physical meaning
of $O$ in the CI parametrization by establishing a relationship
between $O$ and the observable quantities in a generic type-I
seesaw model without any special assumptions. Contrary to the
naive observation, we find that $O$ depends not only on the
neutrino mixing matrix $V$ but also on the matrix responsible for
the charged-current interactions of heavy neutrinos $N^{}_i$. The
latter, which has clear physical meaning and is denoted as $R$,
governs the strength of CP violation in $V$ and that in
leptogenesis. After a detailed analysis of the correlation between
$R$ and $V$, we draw a general conclusion that both unflavored
leptogenesis and flavored leptogenesis have no direct connection
with low-energy CP violation.

\vspace{0.4cm}

\framebox{\large\bf 2} ~ After spontaneous $SU(2)^{}_{\rm L}
\times U(1)^{}_{\rm Y} \rightarrow U(1)^{}_{\rm em}$ symmetry
breaking, the mass terms in Eq. (1) turn out to be
\begin{eqnarray}
-{\cal L}^\prime_{\rm mass} \; = \; \overline{E^{~}_{\rm L}}
M^{~}_l E^{~}_{\rm R} + \frac{1}{2} ~ \overline{\left( \nu^{}_{\rm
L} ~N^c_{\rm R}\right)} ~ \left( \matrix{ {\bf 0} & M^{}_{\rm D}
\cr M^T_{\rm D} & M^{}_{\rm R}}\right) \left( \matrix{ \nu^c_{\rm
L} \cr N^{}_{\rm R}}\right) + {\rm h.c.} \; ,
\end{eqnarray}
where $E$ and $\nu^{~}_{\rm L}$ represent the column vectors of
$(e, \mu, \tau)$ and $(\nu^{~}_e, \nu^{~}_\mu,
\nu^{~}_\tau)^{~}_{\rm L}$, respectively. The overall $6\times 6$
neutrino mass matrix in Eq. (6) can be diagonalized by a unitary
transformation:
\begin{eqnarray}
\left(\matrix{V & R \cr S & U}\right)^\dagger \left( \matrix{ {\bf
0} & M^{}_{\rm D} \cr M^T_{\rm D} & M^{}_{\rm R}}\right)
\left(\matrix{V & R \cr S & U}\right)^*  = \left( \matrix{
\widehat{M}^{}_\nu & {\bf 0} \cr {\bf 0} & \widehat{M}^{}_N}\right)
\; ,
\end{eqnarray}
where $\widehat{M}^{}_\nu = {\rm Diag}\{m^{}_1, m^{}_2, m^{}_3\}$
and $\widehat{M}^{}_N = {\rm Diag}\{M^{}_1, M^{}_2, M^{}_3\}$ have
been defined before. After this diagonalization, the flavor states
of light neutrinos ($\nu^{}_\alpha$ for $\alpha = e, \mu, \tau$) can
be expressed in terms of the mass states of light and heavy
neutrinos ($\nu^{}_i$ and $N^{}_i$ for $i=1, 2, 3$), and thus the
standard charged-current interactions between $\nu^{}_\alpha$ and
$\alpha$ (for $\alpha = e, \mu, \tau$) can be written as
\begin{eqnarray}
-{\cal L}^{}_{\rm cc} \; = \; \frac{g}{\sqrt{2}} ~
\overline{\left(e~~ \mu~~ \tau\right)^{}_{\rm L}} ~\gamma^\mu
\left[ V \left( \matrix{\nu^{}_1 \cr \nu^{}_2 \cr \nu^{}_3}
\right)^{}_{\rm L} + R \left( \matrix{N^{}_1 \cr N^{}_2 \cr
N^{}_3} \right)^{}_{\rm L} \right] W^-_\mu + {\rm h.c.} \;
\end{eqnarray}
in the basis of mass states. So $V$ is just the neutrino mixing
matrix responsible for neutrino oscillations, while $R$ describes
the strength of charged-current interactions between $(e, \mu,
\tau)$ and $(N^{}_1, N^{}_2, N^{}_3)$. $V$ and $R$ are correlated
with each other through $VV^\dagger + RR^\dagger = {\bf 1}$. Hence
$V$ itself is not exactly unitary in the type-I seesaw mechanism and
its deviation from unitarity is simply characterized by
non-vanishing $R$.

Because both $V$ and $R$ are well-defined in Eq. (8), they can be
used to understand the physical meaning of $O$ in the CI
parametrization. To do so, we first derive the seesaw relation
from Eq. (7). The latter yields
\begin{equation}
V \widehat{M}^{}_\nu V^T + R \widehat{M}^{}_N R^T \; = \; {\bf 0}
\; , ~~~
\end{equation}
and
\begin{equation}
S \widehat{M}^{}_\nu S^T + U \widehat{M}^{}_N U^T \; =\; M^{}_{\rm
R} \; .
\end{equation}
If Eq. (7) is rewritten as
\begin{eqnarray}
\left( \matrix{ {\bf 0} & M^{}_{\rm D} \cr M^T_{\rm D} & M^{}_{\rm
R}}\right) \left(\matrix{V & R \cr S & U}\right)^* = \left(\matrix{V
& R \cr S & U}\right) \left( \matrix{ \widehat{M}^{}_\nu & {\bf 0}
\cr {\bf 0} & \widehat{M}^{}_N}\right) \; ,
\end{eqnarray}
we can directly obtain the exact results
\begin{equation}
R \; =\; M^{}_{\rm D} U^* \widehat{M}^{-1}_N \; ,
\end{equation}
and
\begin{equation}
S^* \; =\; M^{-1}_{\rm D} V \widehat{M}^{}_\nu \; .
\end{equation}
Let us substitute Eqs. (12) and (13) into Eqs. (9) and (10),
respectively. Then we arrive at
\begin{equation}
V \widehat{M}^{}_\nu V^T \; =\; -M^{}_{\rm D} \left( U^*
\widehat{M}^{-1}_N U^\dagger \right) M^T_{\rm D} \; ,
\end{equation}
and
\begin{equation}
M^{}_{\rm R} \; =\; U\widehat{M}^{}_N U^T + \left(M^{-1}_{\rm
D}\right)^* V^* \widehat{M}^3_\nu V^\dagger \left(M^{-1}_{\rm
D}\right)^\dagger \; \approx \; U\widehat{M}^{}_N U^T \; .
\end{equation}
The excellent approximation made in Eq. (15) implies that $U$ is
essentially unitary. Taking $U$ to be unitary and combining Eqs.
(14) and (15), we obtain
\begin{equation}
M^{}_\nu \; \equiv \; V \widehat{M}^{}_\nu V^T \; \approx \;
-M^{}_{\rm D} M^{-1}_{\rm R} M^T_{\rm D} \; ,
\end{equation}
where $V$ is also unitary in this approximation. Eq. (16) reproduces
the seesaw formula given in Eq. (2). It is obvious that $R \sim S
\sim {\cal O}(M^{}_{\rm D}/M^{}_{\rm R})$ holds, and thus the seesaw
relation actually holds up to the accuracy of ${\cal O}(R^2)$
\cite{Zhou}.

Now we look at the orthogonal matrix $O$ in the CI parametrization.
Given the basis where $M^{}_{\rm R}$ is diagonal, real and positive,
Eq. (15) implies that $M^{}_{\rm R} \approx \widehat{M}^{}_N$ and $U
\approx {\bf 1}$ are very good approximations. In this case, we get
$M^{}_{\rm D} \approx R \widehat{M}^{}_N$ from Eq. (12).
Substituting this relation into Eq. (3), we obtain
\begin{equation}
O \; \approx \; -i \sqrt{\widehat{M}^{-1}_\nu} ~ V^\dagger M^{}_{\rm
D} \sqrt{\widehat{M}^{-1}_N} \;\; \approx \; -i
\sqrt{\widehat{M}^{-1}_\nu} ~ V^\dagger R \sqrt{\widehat{M}^{}_N}
\;\; ,
\end{equation}
which shows that $O$ is definitely dependent on $V$. It is worth
remarking that both $V$ and $R$, which are respectively associated
with the charged-current interactions of light and heavy Majorana
neutrinos, have clear physical meaning. Hence it seems improper to
draw the conclusion from Eq. (5) that unflavored leptogenesis is
independent of low-energy neutrino mixing and CP violation described
by $V$. If Eq. (17) is substituted into Eq. (5), however, we shall
arrive at a much simpler expression
\begin{equation}
M^\dagger_{\rm D} M^{}_{\rm D} \; \approx \; \widehat{M}^{}_N
R^\dagger R \widehat{M}^{}_N \; .
\end{equation}
This result is actually straightforward, just because of $M^{}_{\rm
D} \approx R \widehat{M}^{}_N$. It apparently has nothing to do with
$V$. So the question becomes whether unflavored leptogenesis depends
on $V$ through $R$. We have known that $V$ is correlated with $R$
via the exact seesaw relation in Eq. (9) and the normalization
condition $VV^\dagger + RR^\dagger = {\bf 1}$. To see this
correlation more clearly, one has to adopt an explicit and
self-consistent parametrization of $V$ and $R$.

\vspace{0.4cm}

\framebox{\large\bf 3} ~ Here we make use of the parametrization of
$V \equiv A V^{}_0$ and $R$ advocated in Ref. \cite{Xing08}:
\begin{equation}
V^{}_0 = \left( \matrix{ c^{}_{12} c^{}_{13} & \hat{s}^*_{12}
c^{}_{13} & \hat{s}^*_{13} \cr -\hat{s}^{}_{12} c^{}_{23} -
c^{}_{12} \hat{s}^{}_{13} \hat{s}^*_{23} & c^{}_{12} c^{}_{23} -
\hat{s}^*_{12} \hat{s}^{}_{13} \hat{s}^*_{23} & c^{}_{13}
\hat{s}^*_{23} \cr \hat{s}^{}_{12} \hat{s}^{}_{23} - c^{}_{12}
\hat{s}^{}_{13} c^{}_{23} & -c^{}_{12} \hat{s}^{}_{23} -
\hat{s}^*_{12} \hat{s}^{}_{13} c^{}_{23} & c^{}_{13} c^{}_{23}
\cr} \right) \; ,
\end{equation}
and
\begin{eqnarray}
A & = & \left( \matrix{ c^{}_{14} c^{}_{15} c^{}_{16} & 0 & 0
\cr\cr
\begin{array}{l} -c^{}_{14} c^{}_{15} \hat{s}^{}_{16} \hat{s}^*_{26} -
c^{}_{14} \hat{s}^{}_{15} \hat{s}^*_{25} c^{}_{26} \\
-\hat{s}^{}_{14} \hat{s}^*_{24} c^{}_{25} c^{}_{26} \end{array} &
c^{}_{24} c^{}_{25} c^{}_{26} & 0 \cr\cr
\begin{array}{l} -c^{}_{14} c^{}_{15} \hat{s}^{}_{16} c^{}_{26} \hat{s}^*_{36}
+ c^{}_{14} \hat{s}^{}_{15} \hat{s}^*_{25} \hat{s}^{}_{26} \hat{s}^*_{36} \\
- c^{}_{14} \hat{s}^{}_{15} c^{}_{25} \hat{s}^*_{35} c^{}_{36} +
\hat{s}^{}_{14} \hat{s}^*_{24} c^{}_{25} \hat{s}^{}_{26}
\hat{s}^*_{36} \\
+ \hat{s}^{}_{14} \hat{s}^*_{24} \hat{s}^{}_{25} \hat{s}^*_{35}
c^{}_{36} - \hat{s}^{}_{14} c^{}_{24} \hat{s}^*_{34} c^{}_{35}
c^{}_{36} \end{array} &
\begin{array}{l} -c^{}_{24} c^{}_{25} \hat{s}^{}_{26} \hat{s}^*_{36} -
c^{}_{24} \hat{s}^{}_{25} \hat{s}^*_{35} c^{}_{36} \\
-\hat{s}^{}_{24} \hat{s}^*_{34} c^{}_{35} c^{}_{36} \end{array} &
c^{}_{34} c^{}_{35} c^{}_{36} \cr} \right) \; , \nonumber \\
\nonumber \\
R & = & \left( \matrix{ \hat{s}^*_{14} c^{}_{15}
c^{}_{16} & \hat{s}^*_{15} c^{}_{16} & \hat{s}^*_{16} \cr\cr
\begin{array}{l} -\hat{s}^*_{14} c^{}_{15} \hat{s}^{}_{16} \hat{s}^*_{26} -
\hat{s}^*_{14} \hat{s}^{}_{15} \hat{s}^*_{25} c^{}_{26} \\
+ c^{}_{14} \hat{s}^*_{24} c^{}_{25} c^{}_{26} \end{array} & -
\hat{s}^*_{15} \hat{s}^{}_{16} \hat{s}^*_{26} + c^{}_{15}
\hat{s}^*_{25} c^{}_{26} & c^{}_{16} \hat{s}^*_{26} \cr\cr
\begin{array}{l} -\hat{s}^*_{14} c^{}_{15} \hat{s}^{}_{16} c^{}_{26} \hat{s}^*_{36}
+ \hat{s}^*_{14} \hat{s}^{}_{15} \hat{s}^*_{25} \hat{s}^{}_{26} \hat{s}^*_{36} \\
- \hat{s}^*_{14} \hat{s}^{}_{15} c^{}_{25} \hat{s}^*_{35}
c^{}_{36} - c^{}_{14} \hat{s}^*_{24} c^{}_{25} \hat{s}^{}_{26}
\hat{s}^*_{36} \\
- c^{}_{14} \hat{s}^*_{24} \hat{s}^{}_{25} \hat{s}^*_{35}
c^{}_{36} + c^{}_{14} c^{}_{24} \hat{s}^*_{34} c^{}_{35} c^{}_{36}
\end{array} &
\begin{array}{l} -\hat{s}^*_{15} \hat{s}^{}_{16} c^{}_{26} \hat{s}^*_{36} -
c^{}_{15} \hat{s}^*_{25} \hat{s}^{}_{26} \hat{s}^*_{36} \\
+c^{}_{15} c^{}_{25} \hat{s}^*_{35} c^{}_{36} \end{array} &
c^{}_{16} c^{}_{26} \hat{s}^*_{36} \cr} \right) \; ,
\end{eqnarray}
where $c^{}_{ij} \equiv \cos\theta^{}_{ij}$ and $\hat{s}^{}_{ij}
\equiv e^{i\delta^{}_{ij}} \sin\theta^{}_{ij}$ with $\theta^{}_{ij}$
and $\delta^{}_{ij}$ (for $1 \leq i < j \leq 6$) being rotation
angles and phase angles, respectively. One can see that $V^{}_0$ is
just the standard parametrization of the {\it unitary} neutrino
mixing matrix (up to some proper phase rearrangements) \cite{PDG},
and thus non-vanishing $A$ signifies the non-unitarity of $V$. One
can also see that $A$ and $R$ involve the same parameters: nine
rotation angles and nine phase angles
\footnote{Note that none of the phases of $R$ (or $A$) can be
rotated away by redefining the phases of three charged-lepton
fields, because such a phase redefinition will also affect the
phases of $A$ (or $R$), as one can easily see from Eq. (8).}.
If all of them are switched off, we shall be left with $R = {\bf
0}$ and $A = {\bf 1}$. In view of the fact that the unitarity
violation of $V$ must be very small effects (at most at the
percent level as constrained by current experimental data on
neutrino oscillations, rare lepton-flavor-violating or
lepton-number-violating processes and precision electroweak tests
\cite{Antusch}), one may treat $A$ as a perturbation to $V^{}_0$.
The smallness of $\theta^{}_{ij}$ (for $i=1,2,3$ and $j=4,5,6$)
allows us to make the following excellent approximations:
\begin{eqnarray}
A & = & {\bf 1} - \left( \matrix{ \frac{1}{2} \left( s^2_{14} +
s^2_{15} + s^2_{16} \right) & 0 & 0 \cr \hat{s}^{}_{14}
\hat{s}^*_{24} + \hat{s}^{}_{15} \hat{s}^*_{25} + \hat{s}^{}_{16}
\hat{s}^*_{26} & \frac{1}{2} \left( s^2_{24} + s^2_{25} + s^2_{26}
\right) & 0 \cr \hat{s}^{}_{14} \hat{s}^*_{34} + \hat{s}^{}_{15}
\hat{s}^*_{35} + \hat{s}^{}_{16} \hat{s}^*_{36} & \hat{s}^{}_{24}
\hat{s}^*_{34} + \hat{s}^{}_{25} \hat{s}^*_{35} + \hat{s}^{}_{26}
\hat{s}^*_{36} & \frac{1}{2} \left( s^2_{34} + s^2_{35} + s^2_{36}
\right) \cr} \right) + {\cal O}(s^4_{ij}) \; , \nonumber \\
\nonumber \\
R & = & {\bf 0} + \left( \matrix{ \hat{s}^*_{14} &
\hat{s}^*_{15} & \hat{s}^*_{16} \cr \hat{s}^*_{24} &
\hat{s}^*_{25} & \hat{s}^*_{26} \cr \hat{s}^*_{34} &
\hat{s}^*_{35} & \hat{s}^*_{36} \cr} \right) + {\cal O}(s^3_{ij})
\;
\end{eqnarray}
with $s^{}_{ij} \equiv \sin\theta^{}_{ij}$ being real. Note that the
approximation made in Eq. (15) is equivalent to $A \approx {\bf 1}$,
leading to unitary $V$ and $U$. One may therefore take $V \approx
V^{}_0$ when applying the approximate seesaw relation in Eq. (2) or
(16) to the phenomenology of neutrino mixing and leptogenesis. In
this case, Eq. (9) is simplified to
\begin{equation}
V^{}_0 \widehat{M}^{}_\nu V^T_0 \approx - R \widehat{M}^{}_N R^T \;
.
\end{equation}
The total number of free parameters in $\widehat{M}^{}_\nu$,
$\widehat{M}^{}_N$, $V$ (or $V^{}_0$) and $R$ is thirty (six masses,
twelve mixing angles and twelve CP-violating phases). But either Eq.
(9) or Eq. (22) can give twelve real constraint conditions. Hence we
are left with eighteen independent parameters in the type-I seesaw
mechanism.

Given the approximate expression of $R$ in Eq. (21), it is
straightforward to obtain
\begin{eqnarray}
{\rm Im} \left( R \widehat{M}^{}_N R^T \right)^{}_{ij} & = & -M^{}_1
s^{}_{i4} s^{}_{j4} \sin \left( \delta^{}_{i4} +
\delta^{}_{j4} \right) \nonumber \\
&& - M^{}_2 s^{}_{i5} s^{}_{j5} \sin \left( \delta^{}_{i5} +
\delta^{}_{j5} \right) \nonumber \\
&& - M^{}_3 s^{}_{i6} s^{}_{j6} \sin \left( \delta^{}_{i6} +
\delta^{}_{j6} \right) \; ,
\end{eqnarray}
where $1\leq i < j \leq 3$. In comparison, Eqs. (4) and (18) tell us
that the CP-violating asymmetries $\varepsilon^{}_i$ (for $i = 1, 2,
3$) in unflavored leptogenesis are associated with
\begin{eqnarray}
{\rm Im} \left( R^\dagger R \right)^{}_{12} & = & \sum^3_{i=1}
s^{}_{i4} s^{}_{i5} \sin \left( \delta^{}_{i4} - \delta^{}_{i5}
\right) \; , \nonumber \\
{\rm Im} \left( R^\dagger R \right)^{}_{13} & = & \sum^3_{i=1}
s^{}_{i4} s^{}_{i6} \sin \left( \delta^{}_{i4} - \delta^{}_{i6}
\right) \; , \nonumber \\
{\rm Im} \left( R^\dagger R \right)^{}_{23} & = & \sum^3_{i=1}
s^{}_{i5} s^{}_{i6} \sin \left( \delta^{}_{i5} - \delta^{}_{i6}
\right) \; .
\end{eqnarray}
We see that there are in general nine independent phase
combinations in Eq. (23), while there are only six independent
phase combinations in Eq. (24). It is possible to acquire ${\rm
Im} ( R \widehat{M}^{}_N R^T) = {\bf 0}$ by fine-tuning the free
parameters in Eq. (23), such that ${\rm Im} ( V^{}_0
\widehat{M}^{}_\nu V^T_0) \approx {\bf 0}$ holds (i.e., the
neutrino mixing matrix $V^{}_0$ is real) as one can see from Eq.
(22). In this special case, there is no low-energy CP violation
but viable unflavored leptogenesis is likely to take place. To
achieve a direct connection between the CP-violating phases of
$V^{}_0$ and the CP-violating asymmetries $\varepsilon^{}_i$, one
should switch off as many phases of $R$ as possible. Such a
treatment can be realized in some specific type-I seesaw models
\cite{Review}, in which the texture of $Y^{}_\nu$ (or $M^{}_{\rm
D}$) might get constrained from a certain flavor symmetry in the
basis of $M^{}_{\rm R} = \widehat{M}^{}_N$. But our general
conclusion is that there is only indirect connection between
unflavored leptogenesis and low-energy observables.

\vspace{0.4cm}

\framebox{\large\bf 4} ~ The same conclusion as obtained above is
true for {\it flavored} leptogenesis. When the mass of the
lightest heavy Majorana neutrino is lower than about $10^{12}$
GeV, flavor-dependent effects matter in leptogenesis
\cite{Barbieri} and have to be carefully handled \cite{Flavor}. In
this case, the CP-violating asymmetries $\varepsilon^{}_{i\alpha}$
between $N^{}_i \to l^{}_\alpha + H^c$ and $N^{}_i \to l^c_\alpha
+ H$ decays (for $i=1,2,3$ and $\alpha =e, \mu, \tau$) depend on
the phases of $M^{}_{\rm D}$ (or $Y^{}_\nu$) in the following way
\cite{Xiong}:
\begin{eqnarray}
\varepsilon^{}_{i\alpha} & = & \frac{1}{8\pi v^2} \sum_{j\neq i}
\left\{ {\cal F}(x^{}_{ij}) ~ \frac{{\rm Im} \left[ (M^\dagger_{\rm
D} M^{}_{\rm D})^{}_{ij} (M^*_{\rm D})^{}_{\alpha i} (M^{}_{\rm
D})^{}_{\alpha j} \right]}{|(M^{}_{\rm D})^{}_{\alpha i}|^2} \right
. \nonumber \\
&& \left . + ~ \frac{1}{1 - x^{}_{ij}} \cdot \frac{{\rm Im} \left[
(M^\dagger_{\rm D} M^{}_{\rm D})^{}_{ji} (M^*_{\rm D})^{}_{\alpha i}
(M^{}_{\rm D})^{}_{\alpha j} \right]}{|(M^{}_{\rm D})^{}_{\alpha
i}|^2} \right\} \; ,
\end{eqnarray}
where the loop function ${\cal F}(x^{}_{ij})$ with $x^{}_{ij} \equiv
M^2_j/M^2_i$ has been given below Eq. (4). Taking account of
$M^{}_{\rm D} \approx R \widehat{M}^{}_N$, we find
\begin{eqnarray}
{\rm Im} \left[ (M^\dagger_{\rm D} M^{}_{\rm D})^{}_{ij} (M^*_{\rm
D})^{}_{\alpha i} (M^{}_{\rm D})^{}_{\alpha j} \right] & \approx &
M^2_i M^2_j ~{\rm Im} \left[ (R^\dagger R)^{}_{ij} R^*_{\alpha i}
R^{}_{\alpha j} \right] \; , \nonumber \\
{\rm Im} \left[ (M^\dagger_{\rm D} M^{}_{\rm D})^{}_{ji} (M^*_{\rm
D})^{}_{\alpha i} (M^{}_{\rm D})^{}_{\alpha j} \right] & \approx &
M^2_i M^2_j ~{\rm Im} \left[ (R^\dagger R)^{*}_{ij} R^*_{\alpha i}
R^{}_{\alpha j} \right] \; .
\end{eqnarray}
It has been shown in Eq. (24) that the quantities $(R^\dagger
R)^{}_{ij}$ (for $i\neq j$) rely on six independent phase
combinations of $R$. On the other hand, it is easy to check that
the quantities $R^*_{\alpha i} R^{}_{\alpha j}$ (for $\alpha = e,
\mu, \tau$ and $i\neq j$) depend on the same phase combinations.
Hence non-vanishing $\varepsilon^{}_{i\alpha}$ in Eq. (25) and
$\varepsilon^{}_i$ in Eq. (4) originate from the same source of CP
violation, no matter whether there are flavor effects or not. This
point keeps unchanged even if resonant leptogenesis \cite{Flavor}
is taken into account.

If the CI parametrization in Eq. (3) is applied to the description
of flavored leptogenesis, then $V$ will show up in the expression
of $\varepsilon^{}_{i\alpha}$. The reason is simply that the
elements of $V$ cannot cancel out in $(M^*_{\rm D})^{}_{\alpha i}
(M^{}_{\rm D})^{}_{\alpha j}$, although they can cancel out in
$(M^\dagger_{\rm D} M^{}_{\rm D})^{}_{ij}$. This observation has
been used by a number of authors to support the argument that
viable flavored leptogenesis may result from $V$ even in the case
of $O$ being a real orthogonal matrix (see, e.g., Refs.
\cite{Petcov,Buras,Petcov2,Branco}). Such an argument is certainly
not wrong, but it is not profound either \cite{Davidson}. In view
of Eq. (17), we find that $O$ can be real only when nontrivial
CP-violating phases in $V$ and $R$ delicately combine to make
$V^\dagger R$ purely imaginary. This extremely special case means
nothing but a very special correlation between $V$ and $R$. While
one may argue that flavored leptogenesis is linked to the neutrino
mixing matrix $V$ in this contrived case, one should keep in mind
that both $\varepsilon^{}_{i\alpha}$ and the CP-violating phases
of $V$ actually originate from $R$ and their direct connection can
only be established when some (or most) of the phase parameters of
$R$ are switched off. In general, however, ``there is no
correlation between successful leptogenesis and the low-energy CP
phase" \cite{Davidson}
\footnote{This conclusion was drawn in Ref. \cite{Davidson} from a
very detailed analysis of the sensitivity of leptogenesis to the
neutrino mixing matrix $V$ by using the CI parametrization and
allowing the elements of $O$ to take arbitrary values in the
parameter space. Here we arrive at the same conclusion by
clarifying the physical meaning of $O$ in an analytic way.}.

\vspace{0.4cm}

\framebox{\large\bf 5} ~ The CI parametrization, in which the
neutrino mixing matrix $V$ and an orthogonal matrix $O$ are
unjustifiedly assumed to be independent of each other, has often
been applied to the phenomenology of neutrino mixing and
leptogenesis in the type-I seesaw mechanism. In the present work,
we have clarified the physical meaning of $O$ by establishing a
relationship between $O$ and the observable quantities in a
generic type-I seesaw model without any special assumptions. We
find that $O$ depends not only on $V$ but also on $R$, the matrix
responsible for the charged-current interactions of heavy Majorana
neutrinos. The CP-violating phases of $R$ govern the strength of
CP violation at low energies and that in leptogenesis. We have
examined the dependence of unflavored or flavored leptogenesis on
$R$ and analyzed the correlation between $R$ and $V$. Our general
conclusion is that both unflavored leptogenesis and flavored
leptogenesis have no direct connection with low-energy CP
violation.

Let us finally give some remarks on $R$, which makes more sense
than $O$ in the analysis of leptogenesis. If the type-I seesaw
mechanism could be realized at the TeV scale, it might be possible
to measure or constrain the mixing angles of $R$ at the Large
Hadron Collider and probe the CP-violating phases of $R$ at a
neutrino factory \cite{Xing}. Because non-vanishing $R$ is a clean
signature of the unitarity violation of $V$, it can actually lead
to rich phenomenology of lepton-flavor-violating and
lepton-number-violating processes. In particular, $R$ bridges a
gap between high-energy neutrino physics (e.g., heavy neutrino
decays and leptogenesis) and low-energy neutrino physics (e.g.,
neutrino mixing and neutrino oscillations).

\vspace{0.6cm}

The author would like to thank S. Zhou for many useful discussions.
This work was supported in part by the National Natural Science
Foundation of China under grant No. 10425522 and No. 10875131.

\end{document}